\documentclass[twocolumn, superscriptaddress, amsmath, secnumarabic, floatfix, amssymb, nobibnotes, aps, pra, reprint]{revtex4-1}
\usepackage{graphicx}
\usepackage{multirow}
\usepackage{xcolor}

\usepackage{microtype}
\usepackage[utf8]{inputenc}
\begin{document}

\title{Patterning-effect-free intensity modulator for secure decoy-state quantum key distribution}

\author{G. L. Roberts}
\affiliation{Toshiba Research Europe Ltd, 208 Cambridge Science Park, Milton Road, Cambridge, CB4 0GZ, United Kingdom}
\affiliation{Cambridge University Engineering Department, 9 JJ Thomson Avenue, Cambridge, CB3 0FA, United Kingdom}
\author{M. Pittaluga}
\affiliation{Toshiba Research Europe Ltd, 208 Cambridge Science Park, Milton Road, Cambridge, CB4 0GZ, United Kingdom}
\affiliation{School of Electronic and Electrical Engineering, University of Leeds, Leeds, LS2 9JT, United Kingdom}
\author{M. Minder}
\affiliation{Toshiba Research Europe Ltd, 208 Cambridge Science Park, Milton Road, Cambridge, CB4 0GZ, United Kingdom}
\affiliation{Cambridge University Engineering Department, 9 JJ Thomson Avenue, Cambridge, CB3 0FA, United Kingdom}
\author{M. Lucamarini}
\affiliation{Toshiba Research Europe Ltd, 208 Cambridge Science Park, Milton Road, Cambridge, CB4 0GZ, United Kingdom}
\author{J. F. Dynes}
\affiliation{Toshiba Research Europe Ltd, 208 Cambridge Science Park, Milton Road, Cambridge, CB4 0GZ, United Kingdom}
\author{Z. L. Yuan}
\email[Corresponding author: ]{zhiliang.yuan@crl.toshiba.co.uk}
\affiliation{Toshiba Research Europe Ltd, 208 Cambridge Science Park, Milton Road, Cambridge, CB4 0GZ, United Kingdom}
\author{A. J. Shields}
\affiliation{Toshiba Research Europe Ltd, 208 Cambridge Science Park, Milton Road, Cambridge, CB4 0GZ, United Kingdom}

\begin{abstract}
Quantum key distribution (QKD) is a technology that allows two users to exchange keys securely.
The decoy state technique enhances the technology, ensuring keys can be shared at high bit rates over long distances with information theoretic security.
However, imperfections in the implementation, known as side-channels, threaten the perfect security of practical QKD protocols.
Intensity modulators are required for high-rate decoy state QKD systems, although these are unstable and can display a side channel where the intensity of a pulse is dependent on the previous pulse.
Here we demonstrate the superior practicality of a tunable extinction ratio Sagnac-based intensity modulator (IM) for practical QKD systems.
The ability to select low extinction ratios, alongside the immunity of Sagnac interferometers to DC drifts, means that random decoy state QKD patterns can be faithfully reproduced with no patterning effects.
The inherent stability of Sagnac interferometers also ensures that the modulator output does not wander over time.
\end{abstract}
\maketitle

Quantum key distribution (QKD) enables the sharing of keys with information theoretic security between two users, Alice and Bob, through standard optical fibers~\cite{Bennett_quantum_1984, Gisin_quantum_2002, Scarani_security_2009}.
The security of QKD is based on the key being transmitted using single photons.
According to the laws of quantum mechanics, a measurement on the photon will disturb it in a way that is observable to the legitimate users.
Despite significant advances in the field of high-rate single photon production, it is not yet at the stage where the sources can be used in practical QKD systems.
Instead, weak coherent pulses are used~\cite{Lutkenhaus_security_2000, Acin_Coherent-pulse_2004}.
These are light pulses that are heavily attenuated so the probability of a pulse containing multiple photons, a scenario that is completely insecure, is vanishingly small.
This technique requires Alice and Bob to estimate the amount of information an eavesdropper, Eve, can obtain, which gives a poor scaling of the secure key rate with distance.
Fortunately, the actual single photon parameters can be bounded using a method known as decoy state QKD~\cite{Hwang_quantum_2003, Lo_decoy_2005, Wang_beating_2005, Lucamarini_efficient_2013, Lim_concise_2014, Lucamarini_security_2015}.
Here, Alice transmits a number of different intensity states, commonly three, allowing the users to accurately determine the maximum amount of information Eve can obtain.
Decoy state QKD has been essential to the development of practical secure communications because it drastically improves the scaling of secure key rate with distance.

Whilst QKD is theoretically secure, imperfections in the actual implementation can create security threats, known as side channels~\cite{Scarani_security_2009, Tamaki_decoy-state_2016}.
One such side channel that has been addressed recently is known as the `patterning effect', and pertains to the production of decoy states in the transmitter~\cite{Yoshino_quantum_2018}.
The finite modulation bandwidth of the system means that the pulse intensity transmitted by Alice can depend on the previous pulse intensity.
This leaks information, allowing Eve to perform a sophisticated attack that reduces the secure key rate when accounted for.
Moreover, the output power of the IMs used to produce the decoy states can vary based on the ambient conditions of the transmitter~\cite{Salvestrini_analysis_2011}.
To counter this, feedback mechanisms are commonly implemented to ensure the IM is faithfully reproducing the correct states~\cite{Fu_Mach-Zehnder:_2013}.
This leads to an increase in system complexity and can mean that time is spent doing stabilisation that could be used to distribute key material.

\begin{figure}[ht!]
	\centering
	\includegraphics[width=\linewidth]{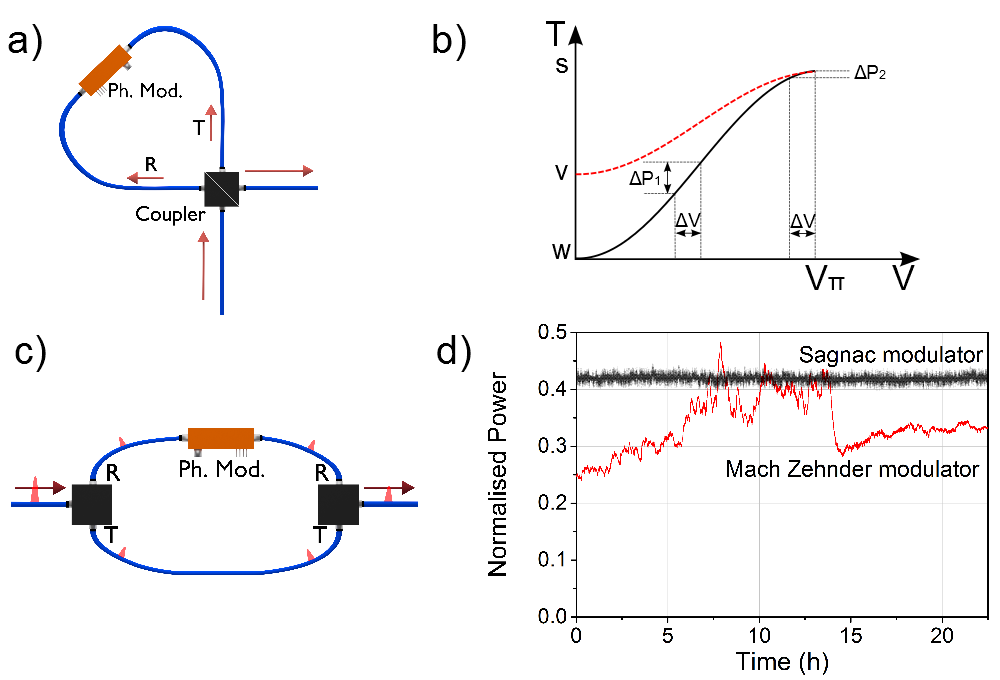}
	\caption{\textbf{Intensity modulation}. \textit{a) Schematic of the Sagnac-based IM with a coupling ratio R:T. b) Transmission (T) with voltage (V) for an interferometer-based IM. Power deviations for identical voltage shifts ($\Delta V$) for signal states (`s') $\left(\Delta P_2 \right)$ and decoy-states (`v') $\left(\Delta P_1 \right)$ are given. The red dotted line shows the output of the proposed low extinction ratio interferometer. c) Schematic of a Mach Zehnder modulator with a coupling ratio R:T. d) The output power from an unmodulated Mach Zehnder IM (red) and the 80:20 Sagnac IM (black) with no feedback. The power is normalized so the maximum power output is unity. }}
    \label{fig:schematics}
\end{figure}

In this paper we describe how a Lithium Niobate (LiNbO$_3$) phase modulator inside a Sagnac interferometer, see Fig.~\ref{fig:schematics}a, can be used as a secure two-level IM for QKD.
By working only at the half-wave voltage of the phase modulator, we ensure that the modulation is at the peak and trough points of the voltage response curve, see Fig.~\ref{fig:schematics}b, while the desired signal-to-decoy intensity ratio is tuned using the coupling ratio of the interfering beamsplitter.
This removes the patterning effect because deviations from these values produce small variations in the output power.
Also, the inherent common-path interference mechanism of the interferometer ensures that there is no DC drift, requiring less time to be devoted to developing feedback mechanisms and further reduces any patterning effects.

The most common IMs for QKD systems are LiNbO$_3$-based Mach-Zehnder interferometers with a phase delay induced in one arm by electrical modulation, as shown in Fig.~\ref{fig:schematics}c.
These are known as Mach-Zehnder modulators (MZMs) and are used to produce the signal (`s'), decoy (`v') and vacuum (`w') states.
Here, the input light field is equally split into two different paths, one of which undergoes a phase shift before they are recombined.
The value of the phase difference defines the intensity of the output, meaning the transmission can be controlled by applying electrical modulations to the phase modulation arm.
The light travels in separate paths, thus each can undergo different phase shifts due to ambient conditions.
This creates a drift in the output intensity that has to be removed using feedback to vary the DC voltage.
MZMs can work up to very high bit rates by using a traveling-wave phase modulator, where a short electrical pulse travels through the device at the same speed as the light pulse~\cite{Wooten_Review_2000}.
Due to the birefringent nature of the LiNbO$_3$ phase modulator crystals, input light must be linearly polarized along one crystal axis.

The patterning effect in these IMs comes from their DC dependence and the sinusoidal response to voltage, as shown in Fig.~\ref{fig:schematics}b (black line).
Small voltage fluctuations, shown as $\Delta V$ in the figure, cause an insignificant variation in the output power, $\Delta P$, for `s' and `w' states.
The `v' state is produced away from these points, however, where small voltage fluctuations can create significant changes in the output power.
At high clock rates, the electrical signal does not have enough recovery time to reach the same base level before the next pulse, effectively changing the DC level.
Also, the mean DC value of the input electrical pattern will vary slightly depending on the random pattern in that section, unless sophisticated encoding schemes are used.
These effects both create voltage fluctuations in a random modulation pattern.

Current commercial IMs are designed to achieve the maximum possible optical extinction ratio.
However, this is not ideal when `v' states with an attenuation of around 6~dB are desired, because the voltage fluctuations will cause large deviations in the power that are dependant on the previous level.
To get around this patterning effect, we propose operating the IM at two levels and designing the device such that the optical extinction ratio can be chosen arbitrarily, as shown by the red dotted line in Fig.~\ref{fig:schematics}b.
This means that regardless of the intensity required, the device can be operated at its half-wave voltage and will faithfully produce the desired intensity levels.
This technique works because the intensity of the light output from one arm of interferometer-based IMs is
\begin{equation}
\label{eq:interference}
  I \propto R^2 + T^2 +2RT\cos(\Delta\phi)
\end{equation}
where $R:T$ is the coupling ratio of the interfering beam splitter(s), $R+T=1$ and $\Delta\phi$ is the difference in phase between the two pulses when they recombine.
This allows us to calculate the optical extinction ratio, expressed in dB as $-10\log_{10}\left(I_{min}/I_{max}\right)$, where $I_{min}$ and $I_{max}$ are obtained from Eq.~\ref{eq:interference} by setting $\Delta\phi$ to 0 or $\pi$, respectively.
The result as a function of $R$ is
\begin{equation}
  ER_{max} = -20\log_{10}(|2R-1|).
\end{equation}
The optical extinction ratio can be chosen to suit the desired application by using a fixed beamsplitter, or it can be tuned with a variable beamsplitter.
The aforementioned commercial IMs are designed to target an infinite splitting ratio, which is obtained for $R=0.5$.
In reality, however, the splitting ratio is never exactly 0.5 and realistic values are between 20 and 30~dB.

Intensity modulators based on Sagnac interferometers work on a similar principle to MZMs, as shown in Fig.~\ref{fig:schematics}a~\cite{Fang_DC_1997}.
The light is again split into two separate paths, denoted as `parallel' and `anti-parallel', in relation to the propagation direction of the modulating electrical traveling wave.
A traveling-wave phase modulator applies a phase shift to the `parallel' wave, meaning the intensity of output light can be controlled.
The major difference from an MZM is that the two light pulses travel through the same length of fiber in a short period of time.
This means that any perturbations to the fiber, or changes to the DC of the phase modulator, affect both pulses equally.
This inherent feature of the modulator means that the device can be very stable and does not require feedback routines~\cite{Wang_practical_2018}.

\begin{figure}[!ht]
	\centering
	\includegraphics[width=0.88\linewidth]{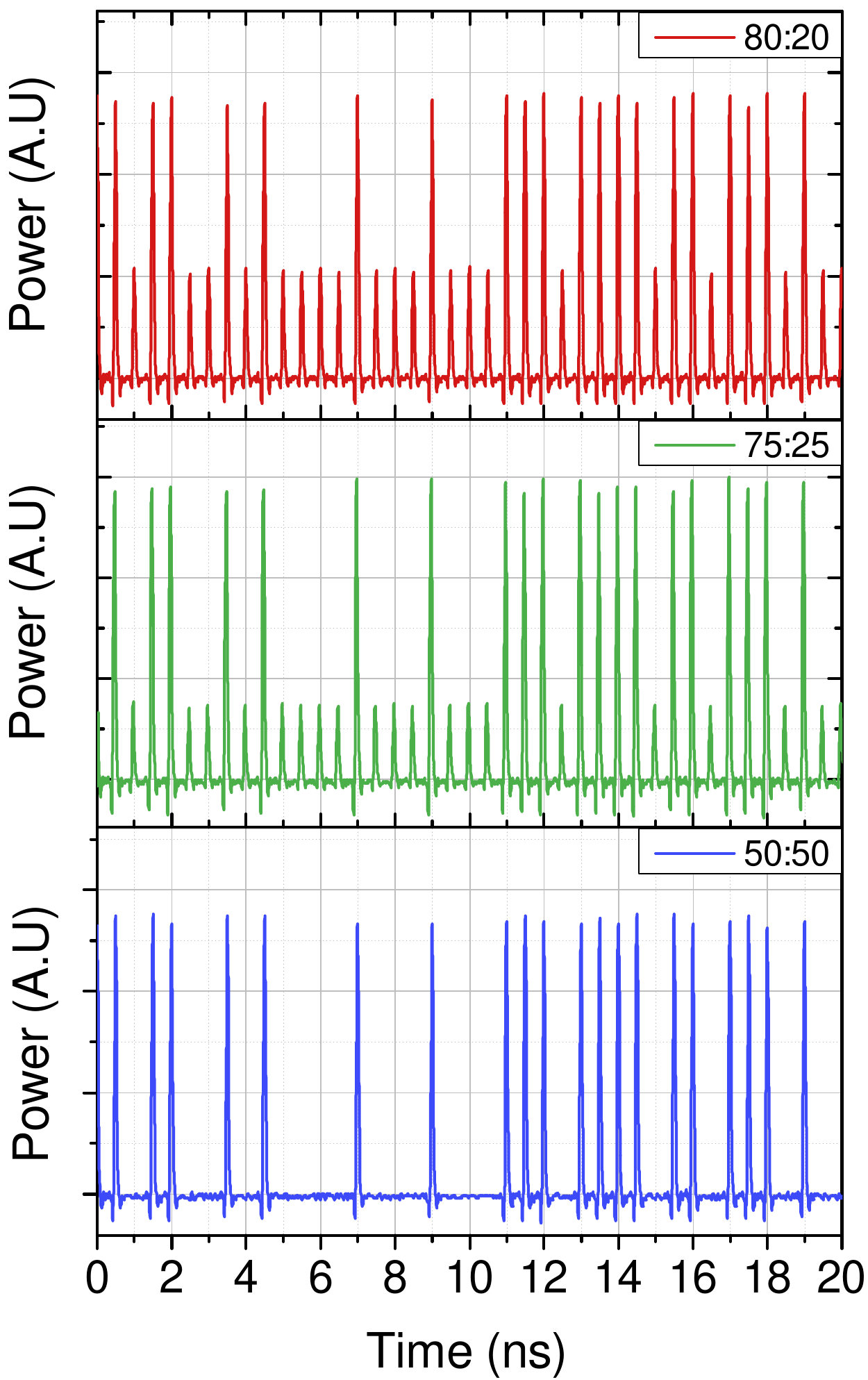}
	\caption{\textbf{Oscilloscope traces}. \textit{Traces at the maximum extinction ratios are shown for a random input pattern for three different beamsplitters. }}
	\label{fig:scopeData}
\end{figure}

To demonstrate their operation, 60~ps light pulses from a 2~GHz gain-switched laser diode are input to the Sagnac IM shown in Fig.~\ref{fig:schematics}a.
The modulator attenuates or blocks light with no electrical input, so 125~ps electrical pulses are input to the phase modulator when a signal pulse is desired.
The electrical pulse delay is tuned so the electrical and light pulses align.
A 1024-bit pseudorandom pattern is generated and the corresponding electrical pattern is applied to the phase modulator.
Short subsets of the resulting outputs are shown in Fig.~\ref{fig:scopeData} for beamsplitters with different coupling ratios.
Three beamsplitters are tested, with nominal splitting ratios of 50:50, 75:25 and 80:20, but realistically providing extinction ratios of 30.48~dB, 5.83~dB and 3.94~dB respectively at their half-wave voltage.

An analysis of the patterning effects for the Sagnac interferometers with a 1024-bit pseudorandom pattern is shown in Table~\ref{tab:Patt}.
Whilst the modulator can be used to produce vacuum states, the lower optical power pulses are referred to as decoy pulses for all coupling ratios.
The transitions to `v' states are not shown for the 50:50 beamsplitter because the photodiode cannot accurately measure such a high optical extinction ratio.
The patterning effects are lower for transitions to `v' states than to `s' states because the `v' states are produced when no modulation is applied.
The patterning effects are negligible for all pulse combinations in all three IMs.
This is especially obvious when compared to the best case scenario of -18.2\% deviation observed by Yoshino~\textit{et al.}~\cite{Yoshino_quantum_2018} when producing decoy states using a commercial MZM at the quadrature point.
The improvement with the Sagnac comes from working only at two levels, as shown in Fig.~\ref{fig:schematics}b, but also on the independence of the modulator on electrical DC drifts.
\begin{table}[ht!]
\centering
\caption{\textbf{Patterning effects}. \textit{Average pulse intensities extracted from a 1024-bit pseudorandom pattern input to Sagnac IMs with three different coupling ratios (CR) when preceded by a decoy pulse (`v') or another signal pulse (`s'). The average `s' pulse intensity is normalized to unity for each beamsplitter. The extinction ratio for each Sagnac (ER) and the deviation from the average intensity are also given.}}
\label{tab:Patt}
\begin{tabular}{@{}cccc@{}}
\toprule
\begin{tabular}[c]{@{}c@{}}CR\\ (ER)\end{tabular}  & Patt. & \begin{tabular}[c]{@{}l@{}}Avg. int.\\ 2nd pulse\end{tabular}  & \begin{tabular}[c]{@{}l@{}}Dev. from\\ avg. (\%)\end{tabular}  \\ \hline
\multicolumn{1}{l}{\multirow{4}{*}{\begin{tabular}[c]{@{}c@{}}80:20\\ (3.94~dB)\end{tabular}}} & s$\rightarrow$s           & $0.999 \pm 0.021$                                 & $0.08$                           \\
\multicolumn{1}{l}{}                       & v$\rightarrow$s            & $1.001 \pm 0.021$                                 & $-0.08$                           \\
\multicolumn{1}{l}{}                       & s$\rightarrow$v           & $0.403 \pm 0.009$                                 & $0.03$                           \\
\multicolumn{1}{l}{}                       & v$\rightarrow$v            & $0.404 \pm 0.009$                                 & $-0.03$                           \\ \hline
\multicolumn{1}{l}{\multirow{4}{*}{\begin{tabular}[c]{@{}c@{}}75:25\\ (5.83~dB)\end{tabular}}} & s$\rightarrow$s           & $0.998 \pm 0.017$                                 & $0.20$                          \\
 \multicolumn{1}{l}{}                       & v$\rightarrow$s             & $1.002\pm 0.017$                                  & $-0.19$                          \\
 \multicolumn{1}{l}{}                       & s$\rightarrow$v           & $0.261 \pm 0.006$                                 & $0.02$                           \\
\multicolumn{1}{l}{}                       & v$\rightarrow$v            & $0.262 \pm 0.006$                                 & $-0.03$                           \\ \hline
 \multicolumn{1}{l}{\multirow{2}{*}{\begin{tabular}[c]{@{}c@{}}50:50\\ (30.48~dB)\end{tabular}}} & s$\rightarrow$s           & $0.999 \pm 0.013$                                 & $0.06$                          \\
 \multicolumn{1}{l}{}                       & v$\rightarrow$s             & $1.001 \pm 0.013$                                  & $-0.06$                          \\ \botrule
\end{tabular}
\end{table}

The difference in stability between a Sagnac IM with an 80:20 beamsplitter and a commercial MZM is shown in Fig.~\ref{fig:schematics}d.
A power meter with a 1~s averaging time is used to measure the output power for modulators with no applied AC.
The Sagnac output shows a Gaussian variation about the mean, with a 1.4~\% standard deviation.
The DC of the MZM is tuned to provide a similar extinction ratio to the Sagnac IM, left for a day to thermally stabilize, and then is left with no feedback.
The output power of the MZM varies unpredictably over a large range of values due to drift, giving a 61.2~\% standard deviation.
The finite variation of the Sagnac modulator is explained by a misalignment of the phase modulator crystal axis causing mixing between orthogonal polarizations.
This is confirmed experimentally by a small observed dependence of the Sagnac output power on the applied DC.
The variation could be reduced further by manufacturing a phase modulator with no misalignment.

With regards to modulator design, the ideal case is where the phase modulator is placed asymmetrically in the Sagnac loop.
When placed in the center, the `parallel' light pulse has an interaction length of the whole phase modulator because of the co-propagating electrical pulse, whereas the `anti-parallel' light is also modulated by the counter-propagating electrical pulse, albeit with a much smaller interaction length.
A carefully designed offset from the center ensures the `anti-parallel' light does not interact with the electrical pulse.
At higher clock rates this interaction is unavoidable, regardless of the system design, leading to a slightly higher half-wave voltage for the Sagnac IM than that of the phase modulator.
If the input clock rate is too high, however, patterning effects will start to emerge because multiple `anti-parallel' light pulses will be modulated by a single counter-propagating electrical pulse.
This limits the maximum clock rate to 3~GHz for ordinary bulk phase modulators with a crystal length of 5~cm, however can be much higher if smaller phase modulators are used~\cite{Spickermann_gaas/algaas_1996}.

With regards to how this device could be implemented in a QKD system, two decoy-state QKD would require two Sagnac IMs to remove the patterning effects.
Fortunately, the modulator stability would mean that this does not add too much complexity to the system.
Even still, using as few components as possible would be ideal.
One way this could be done is with a single decoy-state QKD protocol, which would require just a single IM~\cite{Ma_practical_2005, Rusca_finite-key_2018}.
Another possibility would be to use a directly-modulated injection-locked quantum transmitter~\cite{Yuan_directly_2016, Roberts_modulator-free_2017}.
This has been shown to provide both phase modulation and high extinction ratio intensity modulation.
This transmitter could produce the signal and vacuum states, requiring just a single IM for the decoy states.

We have successfully demonstrated a variable extinction ratio IM to provide the decoy states in QKD systems.
Both high extinction ratios of 30.48~dB and low extinction ratios of 5.83~dB and 3.94~dB have been shown using a 50:50, 75:25 and 80:20 beamsplitter respectively.
A variable beamsplitter could also be used, allowing the decoy state intensity to be accurately tuned.
Selecting a suitable beamsplitter and then working solely at the half-wave voltage of the the DC-independent Sagnac IM ensures that there is no correlation between the produced intensities, which is not the case for IMs operated around the quadrature point with a DC dependence.
We have shown that a Sagnac IM, unlike an MZM, has no temporal drift, meaning feedback mechanisms are not required, thus dramatically simplifying the implementation.

\section*{Acknowledgments}

G. L. R. acknowledges financial support via the EPSRC funded CDT in Integrated Photonic and Electronic Systems, Toshiba Research Europe Limited and The Royal Commission for the Exhibition of 1851.
M. P. acknowledges funding from the European Union’s Horizon 2020 research and innovation programme under the Marie Skłodowska-Curie grant agreement No 675662.
M. M. acknowledges financial support from the EPSRC and Toshiba Research Europe Ltd.

% Bibliography
\bibliography{sample}

\end{document}